\documentstyle[12pt,equation,prl,aps,preprint,tighten,floats,aps,epsf,psfig]{revtex}

\def\be{\begin{equation}}
\def\ee{\end{equation}}
\def\bear{\begin{eqnarray}}
\def\eear{\end{eqnarray}}
\newcommand{\beq}{\begin{eqalignno}}  
\newcommand{\eeq}{\end{eqalignno}}

\def\ncom{\newcommand}
\ncom{\mbf}[1]{\mathbf{#1}}
\ncom{\mca}[1]{\mathcal{#1}}

\begin{document}
\draft

\title{Implications of Supersymmetry Phases for Higgs Boson Signals and 
Limits}

\author{G.L.Kane\footnote{email gkane@umich.edu} and Lian-Tao Wang
\footnote{email: liantaow@umich.edu} \\ 
\it Randall Physics Laboratory, \\ Dept. of Physics, \\ Ann Arbor, MI
48109, USA}

\maketitle

\begin{abstract}
We study the supersymmetry parameter region excluded if no Higgs is
found at LEP, and the region allowed if a Higgs boson is found at LEP.
We describe the full seven parameter structure of the Higgs sector. When
supersymmetry phases are included, $\tan \beta \stackrel{>}{\sim}2$ is
always allowed, and the lower limit 
on $M_{H^1}$ if no signal is found is about $20 \%$ lower
than in the Standard Model and about $10 \%$ lower than in the MSSM with
phases set to $0$, $\pi$. 

\end{abstract}

\newpage

\section{Introduction}

A general supersymmetric theory has a large parameter space. This is not
a major obstacle to study the kinds of signatures to expect because the
parameters split up into sets so that only a limited number affect a
given process. Once there is data on superpartners the observed
patterns
will quickly allow us to reduce the number of parameters, just as
happened for the standard model in the past.

Early in the history of supersymmetry phenomenology, both for
simplicity and in order to learn
qualitatively
how cross section and decays behaved, simplifying assumptions have been
made about the supersymmetry parameters. That was useful. Of course,
results based on those assumptions do not hold in general. One
purpose of the present paper is to demonstrate that this can have
a large effect when non-observation of a signal is used to set limits,
or when (hopefully soon) a signal is analyzed to extract information
about $\tan \beta$ and the supersymmetry Lagrangian. 

The phases of the soft breaking Lagrangian are particularly important.
They can play a major role in CP violation, and could provide the CP
effects \cite{kane:1999ef} that explain the baryon asymmetry in the
universe (which can not  be understood in the Standard Model). They
could have a major impact on
the formulation of how to compactify string theory and break
supersymmetry. That the phases can significantly affect the Higgs sector
was first pointed out in \cite{kane:1998me} , and independently in
\cite{pilaftsis:1998}, and subsequently well studied in
\cite{kane:1998ed},...,\cite{10}. Although these studies showed
that large effects were possible, two interesting questions were not
specifically  
analyzed. Here we show results for those two questions, which are of
considerable interest for the upcoming LEP run and for Fermilab. We
understand that a similar analysis of LEP chargino limits will soon
appear.\cite{11}

The questions are: (a) If no Higgs boson signal is observed, what is the
general lower limit on the lightest Higgs boson mass and the associated
limit on $\tan
\beta$? (b) If a Higgs boson is observed, what region of the
supersymmetry parameters is consistent with the measured mass and
$\sigma \times BR$? In both cases we show that the results differ
significantly if the phases are fixed at $0$, $\pi$ as is usual,
compared to the situation with general phases allowed. 

A complete analysis of the questions can only be done by the
experimenters who include efficiencies. We only examine some special
cases (which we check are typical) in order  to demonstrate the
importance of this analysis. Qualitatively, we find that if a Standard
Model Higgs boson of mass $105$ $GeV$ were excluded, the associated mass
limit for the SUSY case with phases $0$ or $\pi$ would be about $95$
$GeV$, and
with phases fully included would be about $85$ $GeV$. No LEP Higgs
results can exclude values of $\tan \beta$ larger than $2$ (the region
between $1.5$ and $2$ is also probably allowed but only for a
somewhat narrower
parameter 
space). (Even current results do not exclude $\tan \beta = 2$ when
phases are included. ) The results of our analysis are shown in Figures
$1$,
2.


\section{Framework}

\subsection{Higgs Mass Matrix and Mixing}

It is known \cite{pilaftsis:1998} that with one-loop radiative
corrections and non-zero CP-violating phase in the MSSM soft Lagrangian,
the VEVs of the Higgs doublets can not be chosen to be real
simultaneously. Therefore, most generally, we should parameterize the
Higgs doublets to be

\be
H_1=\frac{e^{i\alpha_1}}{\sqrt{2}} \left(\begin{array}{c}
v_1+h_1 + ia_1 \\ h_1^{-}\end{array} \right).
\ee
\be
H_2=\frac{e^{i\alpha_2}}{\sqrt{2}} \left(\begin{array}{c}
h_2^{+} \\ v_2+h_2 + ia_2 \end{array} \right).
\ee

In the MSSM superpotential and the soft Lagrangian, the only places
where the phases $\alpha_1$ and $\alpha_2$ appear and can not be
rotated
away by a redefinition of other matter superfields are the terms $\mu
H_1 H_2$ and $m_3^2 H_1 H_2$. Those are also the terms relevant to the
scalar potential which determines the vacuum expectation values.
therefore in
the scalar potential the phases only enter in the combination $\theta =
\alpha_1
+ \alpha_2$. One can then choose to parameterize the Higgs doublets as

\be
H_1=\frac{1}{\sqrt{2}} \left(\begin{array}{c}
v_1+h_1 + ia_1 \\ h_1^{-}\end{array} \right).
\ee
\be
H_2=\frac{e^{i\theta}}{\sqrt{2}} \left(\begin{array}{c}
h_2^{+} \\ v_2+h_2 + ia_2 \end{array} \right).
\ee

The neutral part of the Higgs doublets are
\be 
H_1^0 = \frac{1} {\sqrt{2}} \left( h_1 + i a_1 \right); \ \ \ \ \
H_2^0 = \frac {{\rm e}^{i \theta}} {\sqrt{2}} \left( h_2 + i a_2
\right).
\ee
Define $\tan \beta \equiv v_2/v_1$ (the ratio of the magnitudes of the 
VEVs). Note that with our choice of
parameterization, we can define $\tan \beta$ to be real without loss of
generality. 

The neutral Higgs potential including one-loop radiative correction is

\beq
V &=\frac{1}{2} m_1^2(h_1^2 + a_1^2) + \frac{1}{2} m_2^2 (h_1^2 + a_1^2)
-
m_3^2[(h_1 h_2 - a_1 a_2) \cos \theta-(h_1 a_2 + h_2 a_1) \sin
\theta] \nonumber \\
 &+ \frac{\bar{g}^2}{8} D^2 + \frac{1}{64\pi^2}Str \left[{\mca{M}}^4
\left(\log \frac{{\mca{M}}^2}{Q^2} - \frac{3}{2}\right) \right]
\eeq 
where we define
\be
D=h_1^2 - h_2^2 + a_2^2 - a_1^2 ; \ \  \  \ \ \bar{g}^2= \frac{g_1^2
+ g_2^2}{4}.
\ee
$Q$ is the renormalization scale. The supertrace is defined by $Str=
\sum_{J}(-1)^{2J+1}(2J+1)$. $\mca{M}$ is the Higgs field dependent mass
matrix of particles.  

We will only consider the contributions from the top/stop loop and also
neglect the contribution from the D-term \cite{demir:1999}. From the
stop mass Lagrangian 

\be
{\cal L} = [M_{\tilde L}^2 + \frac{1}{2}h_t^2((v_2+h_2)^2 + a_2^2)]
{\tilde t_L}^*{\tilde t_L} +h_t \left[ A_t H_2^0  -
\mu^* H_1^{0*} \right] {\tilde t_R}^*{\tilde t_L} + (L
\rightarrow R),
\ee
we find the mass scalar top mass matrix 

\beq
{\cal M}_{\tilde t}^2 &= \mbox{$ \left( \begin{array}{cc}
m^2_{\tilde L} + \frac{1}{2}h_t^2((v_2+h_2)^2 + a_2^2)
&
 h_t \left[ A_t H_2^0  - \mu^* H_1^{0*} \right] \\
h_t^* \left[ A_t^* H^{0*}_2 - \mu  H_1^0  \right] &
m^2_{\tilde R} + \frac{1}{2}h_t^2((v_2+h_2)^2 + a_2^2)
\end{array} \right), $} 
\eeq 
where $h_t$ is the Yukawa coupling between Higgs and the top quark
which is related to the top quark mass by $h_t=\sqrt{2} m_t/v_2$.  From
diagonalizing this matrix, we see the phases will only enter the
radiative corrections in the 
combination $\gamma=\phi_{\mu}+\phi_{A_t}+\theta$, where $\phi_{A_t}$
and $\phi_{\mu}$ are the phases of $A_t$ and $\mu$, respectively (as
they must by reparameterization invariance).

By minimizing the Higgs potential we get 3 equations. Two of them are
obtained by varying the Higgs potential with respect to $h_1$ and $h_2$.
There are also 2 equations coming from varying $a_1$ and $a_2$ which are
actually not independent and yield one nontrivial equation for the phase 
$\theta$ \cite{demir:1999,choi:2000},

\be
\label{theta}
m_3^2 \sin \theta = \frac{3h_t^2}{32 \pi^2} |\mu| |A_t| sin\gamma
f(m_{\tilde{t}_1}^2, m_{\tilde{t}_2}^2),
\ee
where 

\be
f(x_1,x_2)=2x_1\left(\log \frac{x_1}{Q^2} -1 \right) - 2x_2\left(\log
\frac{x_2}{Q^2} -1 \right)
\ee 

As usual, one can single out  the massless Goldstone boson to be eaten
by
the $Z^0$ as

\be
G^0=-\cos \beta a_1 + \sin \beta a_2.
\ee
Now because we have the  CP-violating phase in the Higgs potential, the
so
called pseudoscalar  $A^0= \sin \beta a_1 + \cos \beta a_2$ will mix
with
the other two neutral  scalars $h_1^0$ and $h_2^0$. So in the basis of
$(h_1^0, A^0, h^0_2)$,  we will have a $3 \times 3$ mass matrix. The
mass matrix is real and symmetrical \cite{demir:1999,choi:2000}. 
Therefore, it  can be diagonalized by an orthogonal transformation

\be
{\mbf{U}} {\mbf{M}}^2  {\mbf{U}}^T =
diag(M_{H^1}^2,M_{H^2}^2,M_{H^3}^2),
\ee
where ${\mbf{U}}$ is a  real and orthogonal matrix satisfying
${\mbf{U}}{\mbf{U}}^T={\mbf{1}}$. $H^i$ is the $i$th Higgs mass
eigenstate. The basis of the mass matrix is chosen in such a way that
when CP-violation goes to zero, $H^2 \rightarrow A$, the usual
pseudoscalar, and $H^1$ is the lightest neutral scalar.

Let's pause for a moment here to count the relevant parameters. In
the original Higgs potential, we have 12 parameters,

\beq
& v_2, v_1,  \phi_{\mu}+\phi_{A_t}, \theta=\alpha_1 + \alpha_2,|A_t|, 
\nonumber \\  & |\mu|,M^2_{\tilde{L}},
M^2_{\tilde{R}},b=m_3^2,m^2_{H_1},m^2_{H_2} \quad\mbox{and}\quad  Q
\mbox{(the renormalization scale).}
\eeq

At the same time,  we have the 3 equations obtained by minimizing the
potential. So we can use them to eliminate 3 of the parameters. We
choose to proceed as following 

\begin{enumerate}
\item We use two of the extremization conditions (other than
the equation for phase $\theta$, Eq.~(\ref{theta})) to eliminate
the parameters $m^2_{H_1}$, $m^2_{H_2}$. We use relation $M_Z^2=(g_1^2 +  
 g_2^2)(v_1^2+v_2^2)/4$ and definition $\tan \beta \equiv v_2/v_1$ to
rewrite $v_1$ and $v_2$ in terms of $M_Z^2$ and $\tan \beta$. This also
inputs one piece of data, the value of $M_Z$, so one more parameter is
eliminated. By doing so, we also make sure the electroweak symmetry
breaking occurs appropriately.

\item For any given set of parameters, solve Eq.~(\ref{theta})) for
$\theta$. So effectively this phase is a function of other SUSY
parameters. 

\item  The renormalization scale $Q$ should not be counted as a
parameter; we set it to be $200$  $GeV$. In a more complete analysis,
higher
order
corrections could be used to fix $Q$ so that further loop corrections
are minimized.

\item Then we can regard the Higgs mass matrix as depending on the
following 7 parameters

\be
\tan \beta,  \phi_{\mu}+\phi_{A_t}, |A_t|, |\mu|, M^2_{\tilde{L}},
M^2_{\tilde{R}},b=m_3^2.
\ee
For any given set of the above 7 parameters, we numerically diagonalize
the
mass matrix to get the eigenvalues and mixing matrix elements, $U_{ij}$; 
the latter enters into the cross sections and decay branching ratios.
(If $\tan \beta$ were large even more parameters could enter from
sbottom loops.)

\end{enumerate}

\subsection{The Relevant Lagrangian}

We present here briefly the relevant Lagrangian of Higgs production and
decay. The Feynman rules can be read from them and the calculation of
the
production cross section and branching ratios is straightforward. 

\begin{enumerate}
\item $VV H$. The Lagrangian is the same for both $W$ and $Z$ bosons.
   
\be
{\mca{L}}_{VV H} =  (\cos \beta U_{1i} +\sin \beta U_{3i}) V^{\mu}
V_{\mu} H^i .
\ee

\item For the process $Z \rightarrow H^1 H^2$, we have

\be
{\mca{L}}_{Z H H} = (\sin \beta U_{2i} U_{1j} - \cos \beta U_{2i}
U_{3j}) Z^{\mu} H^i \stackrel{\leftrightarrow}{\partial}_{\mu}
H^j
\ee

\item The $H \rightarrow b\bar{b}$ vertex
   
\be
{\mca{L}}_{b \bar{b} H} =\frac{m_b}{v \cos \beta} \bar{b}
(U_{1i}-i\sin \beta U_{2i} \gamma_5)b H^i
\ee
We get a similar expression for the Higgs decaying into a pair of
leptons. 

\item The $H\rightarrow  c\bar{c}$ vertex

\be
{\mca{L}}_{c \bar{c} H} =\frac{m_c}{v \sin \beta} \bar{c}
(U_{3i}-i\cos \beta U_{2i} \gamma_5)c H^i
\ee
\end{enumerate}

The branching ratio we are interested in for this paper is
$BR=\Gamma(H^i
\rightarrow b\bar{b})/\Gamma_{Total}$ where we approximate
$\Gamma_{Total}$ by

\be
\Gamma_{Total}=\Gamma(H^i \rightarrow b\bar{b}) + \Gamma(H^i
\rightarrow c\bar{c}) + \Gamma(H^i \rightarrow \tau \bar{\tau}).
\ee
If additional decays can occur, they can be incorporated; the
qualitative results will be similar. How one normalizes the
observable branching ratios is detector-dependent, so we make a simple
choice to illustrate the effects.

\section{Numerical Results}

We will address the effects of phase on the following two problems:

\begin{enumerate}
\item Suppose no Higgs boson is found at LEP. Then in practice there is
a limit on $\sigma(e^+ e^- \rightarrow ZH^1)\times BR(H^1 \rightarrow
b\bar{b})$. Given that limit, what is the general lower limit on $M_H^1$
in the full 7-parameter space? We compare this with the case of setting
the phase to $0$ or $\pi$ as normally done.

\item Suppose a Higgs Boson is found. Its mass and $\sigma \times
BR(b\bar{b})$ are measured. What region of the general 7-parameter space
is allowed? We compare the this with the case when the phase is set to
$0$ or $\pi$ as is usually done.  

\end{enumerate}

\subsection{Lower Limit of Higgs Mass and $\tan \beta$}

If we have an experimental limit on the Higgs production $\times$
branching ratio, we can determine a lower limit on the Higgs mass in 
the Standard Model. What we are interested in is: if we use the Minimal
Supersymmetric Standard Model, including {\it
non-zero} CP violating passe, how will the results one would get in the
MSSM without phases change?

To investigate this question, we proceed as follows.

\begin{enumerate}
\item Pick a Higgs mass $M_H^{(SM)}$. Then we use the Standard Model to
calculate the $\sigma_{prod} \times BR_{(SM)}$. Suppose this
calculated $\sigma_{prod} \times BR_{(SM)}$ is in fact the experimental
limit, then $M_H^{(SM)}$ is going to be the predicted lower limit on the
Higgs mass using the Standard Model. 

\item Then we switch to the MSSM. For any given set of parameter we
calculate $\sigma_{prod} \times BR_{(MSSM)}$ and the corresponding
$M_{H^1}$. We scan the parameter space. If we find that for certain
parameters, $\sigma_{prod} \times BR_{(MSSM)}$ $<$ $\sigma_{prod}
\times BR_{(SM)}$ (The experimental limit) and at the same time
$M_{H^1}$ $<$ $M_H^{(SM)}$ (the lower limit predicted by Standard
Model), then we get a potential candidate for a new lower limit of
Higgs mass. 

\item In order to achieve the above requirement, one must find a set of
parameters such that the coupling of $H^1$ to $Z$ is smaller than in the 
Standard Model. However,
this is potentially
dangerous. When the coupling for $ZZH^1$ decreases, the coupling for
process $Z \rightarrow {H}^1 {H}^2$ increases. So we also need to 
make sure that the cross section for $Z \rightarrow H^1
H^2$ times its branching ratio to $bb\bar{b} \bar{b}$ won't get
so large that it would have been seen. This is a complicated
issue related to the $b-$tagging efficiency. Since our purpose is to
illustrate typical effects rather than to settle the issue, we will
make a
crude estimate and require that $\sigma_{Z \rightarrow ZH^1}
\times BR(H_1)$ $>$ $\sigma_{Z \rightarrow H^2 H^1} \times BR(H_1)
\times BR(H_2)/3 $. The factor of $1/3$ is assumed to reflect
efficiencies. 
\end{enumerate}

The seven dimensional parameter space makes it difficult to illustrate
the
most general results. Consequently we simply illustrate with a few
parameters fixed to show the effect of the CP-violating phase. We
emphasize that we do not in any sense argue that the parameters we fix
should be thought to have those actual values. We have checked that the
results we present are typical.   The results
are shown in Figure 1. We see that if we allow a non-zero phase, we get
a
new potential lower limit on the Higgs mass around $85$ $GeV$. And
lower
$\tan \beta $ values become allowed. If the phase is $0$ or $\pi$, the
usual results emerge.

\subsection{MSSM Parameter Region}

First we pick a set of parameters to generate a set of 'fake' data (a
cross section $\times$ branching ratio and a mass). Then we take into
account estimated
experimental errors ($10{\%}$ for cross section and $5{\%}$ for mass) and
find the range of parameters that can generate such a set of data.
Again, to illustrate the effects of phases, it suffices to focus on a
typical 
region of the parameter space. So we will fix $M_{\tilde{L}}$,
$M_{\tilde{R}}$ and $\mu$ at the values used to generate the fake
data and allow  $A_t$, $\tan \beta$ and $m_3$ to vary. We compare
the results from  fixing the phase to be $0$, $\pi$ with the results
allowing the phase to vary. We illustrate the results by showing the
allowed region in the  $\tan \beta$ and $2m_3^2/\sin 2\beta$ plane in
Figure 2; ($2m_3^2/\sin 2\beta$ would be $M_A$ in the absence of CP
violation).

We see the presence of phases allows an  enlargement of the parameter
region and in particular it allows lower values of  $\tan \beta$. 

\section{Conclusion}

If nature is indeed supersymmetric, the soft supersymmetry breaking
Lagrangian is the basic thing to be determined. Masses of mass
eigenstates and cross sections $\times$ branching ratios are measured,
and
have to be converted into the parameters of the Lagrangian. The patterns
in the resulting Lagrangian will help (and may be essential) to
formulate
how to convert 11D supersymmetric M-theory into a 4D theory with broken
supersymmetry, so that M-theory becomes testable in the normal sense.

Extracting the parameters of the Lagrangian must be done with full
generality. In this analysis we have shown that including the
supersymmetry phases (that are usually ignored) can have qualitative
large effects on how both the presence and absence of a signal for a
Higgs boson are interpreted. Similar effects will occur for superpartner
limits. 

The results are in the two figures. Qualitatively, we find that the 
excluded Higgs boson mass is about $20 \%$ lower than it would be in the
Standard
Model and about $10 \%$ lower than it would be in the MSSM without
phases, and that $\tan \beta$ values down to 2 or even lower are always
allowed by LEP results. We have focused on LEP, but a similar analysis
is required for Fermilab and LHC.

\section{Acknowledgments}

We appreciate helpful comments from M.Brhlik, L.Everett, S.Mrenna, and
J.Wells.

\newpage

\begin{figure}[h!]
\centering
\epsfxsize=12.0cm
\hspace*{0in}
\epsffile{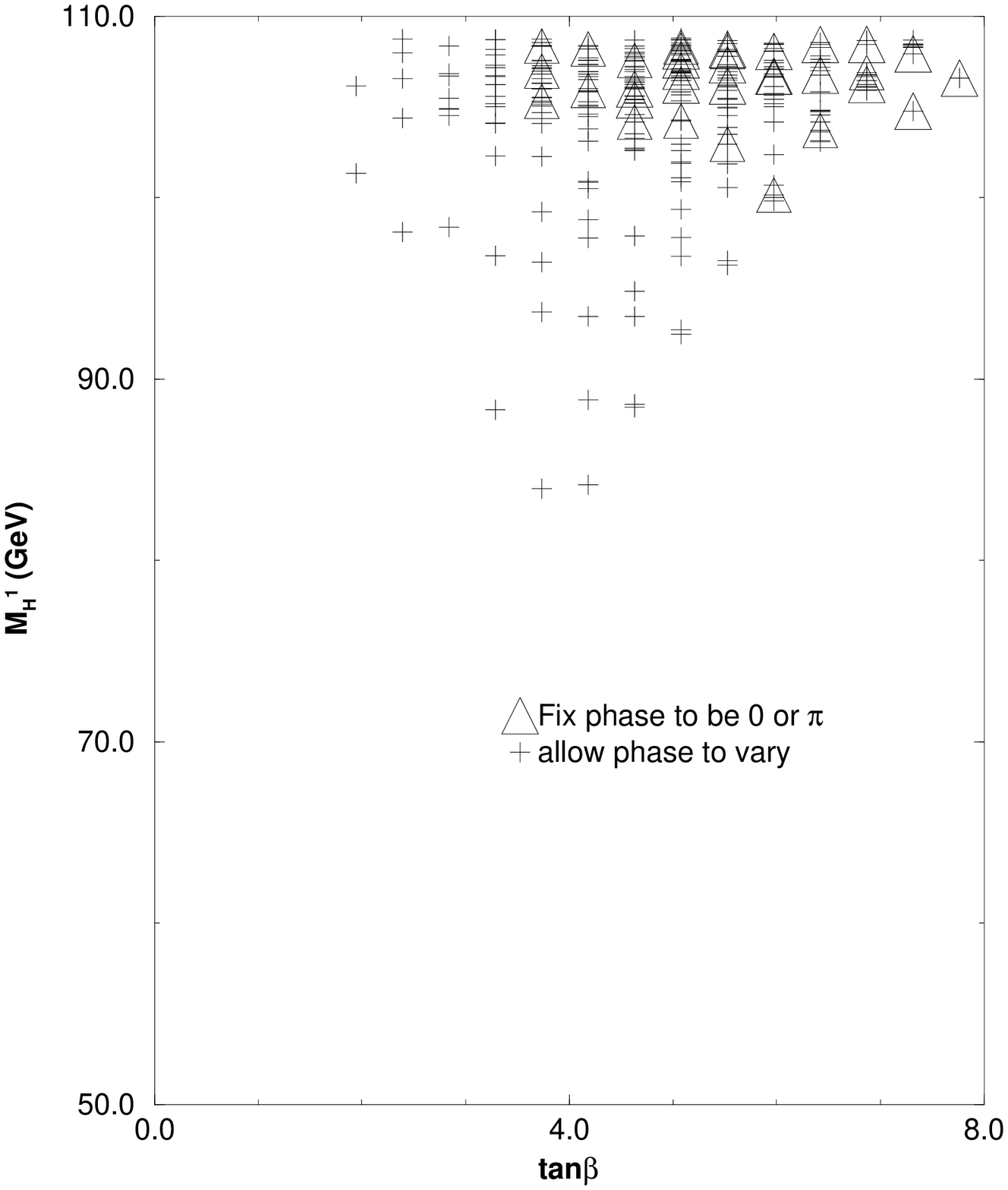}
\bigskip

\caption{The allowed MSSM parameter region if no Higgs has been found. 
We set $M_{\tilde L}=300GeV$, $M_{\tilde R}=250GeV$. We pick two
points in the $(A_t,\mu)$ space: 
$A_t=500GeV$, $\mu=300GeV$ and $A_t=350GeV$, $\mu=250GeV$. We vary
$50GeV < m_3<100GeV$,
$0<\phi_{\mu}+\phi_{A_t}<\pi$
and $2<\tan \beta<10$. We use $M_{H}^{(SM)}=105GeV$. The $\triangle$s
show the region of this plane would be allowed if the phase were fixed
at $0$ or $\pi$, while the $+$ show that a significantly different
region is allowed for the full acceptable range of the phases}

\label{figone}
\end{figure}

\newpage

\begin{figure}[h!]
\centering
\epsfxsize=12.0cm
\hspace*{0in}
\epsffile{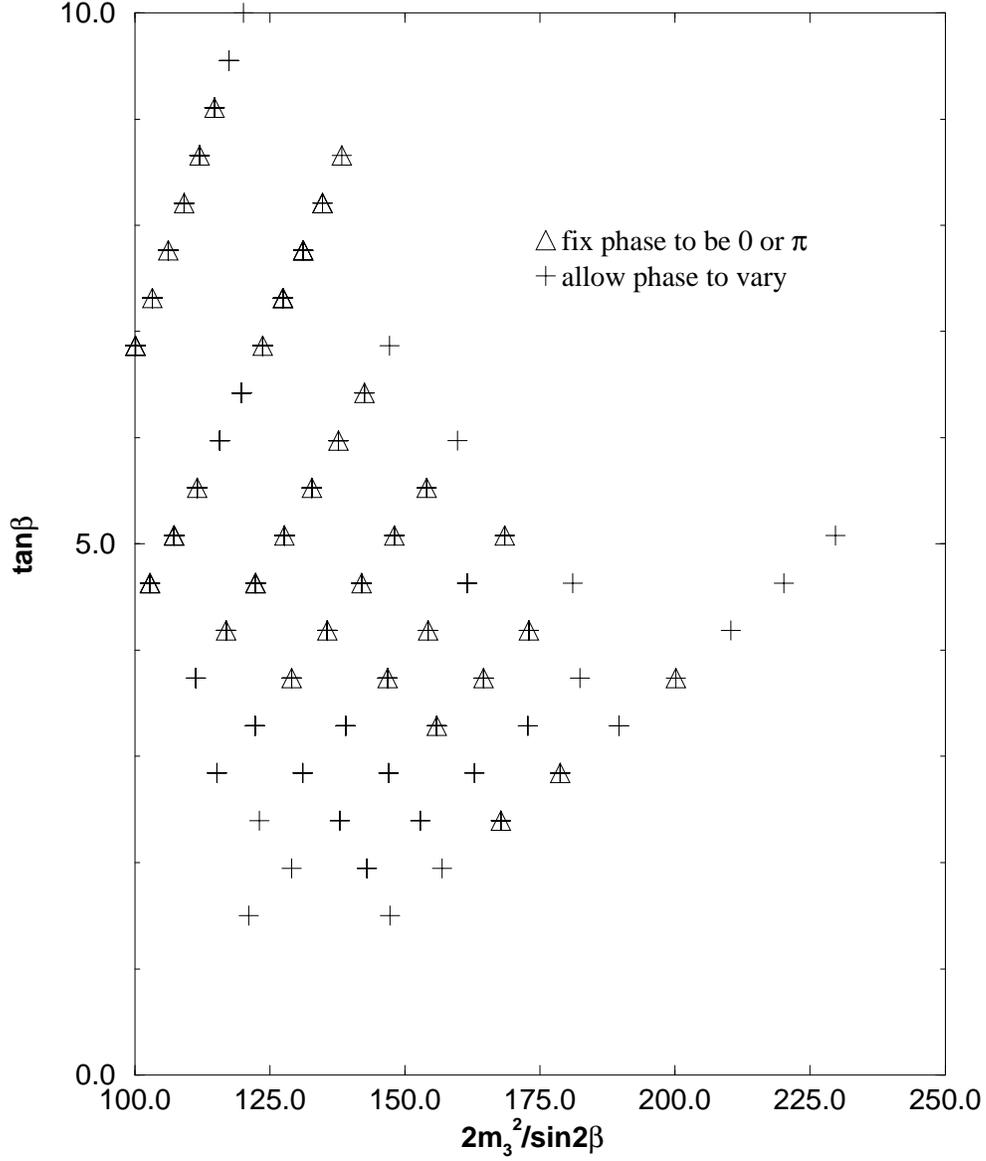}
\bigskip
\caption{The allowed MSSM parameter region if one neutral Higgs has been
found.
Fake data generated by $M_{\tilde L}=300GeV$, $M_{\tilde R}=250GeV$,
$A_t=500GeV$,
$\mu=300GeV$,  $\tan \beta=6.0$, $\phi_{\mu}+\phi_{A_t}=0.0$ and
$m_3=50GeV$. The horizontal axis would be $M_A$ if there were no CP
violation in the Higgs sector. The results show that a larger region of 
$\tan \beta$ and "$M_A$" is allowed if the effects of phases are
included}

\label{figtwo}
\end{figure}

\end{document}